\documentclass{PNASTWO}

\begin{document}

\title{Echo chambers in the age of misinformation}

\author{Michela Del Vicario\affil{1}{Laboratory of Computational Social
    Science, Networks Dept IMT Alti Studi Lucca, 55100 Lucca, Italy}, 
Alessandro Bessi\affil{2}{IUSS Institute for Advanced Study, 27100 Pavia, Italy}, 
Fabiana Zollo\affil{1}{Laboratory of Computational Social Science,
  Networks Dept IMT Alti Studi Lucca, 55100 Lucca, Italy}, 
Fabio Petroni\affil{3}{Sapienza University, Rome, Italy},
Antonio Scala \affil{1},\affil{4}{ ISC-CNR Uos "Sapienza", 00185 Roma, Italy},
Guido Caldarelli \affil{1},\affil{4}{},
H. Eugene Stanley \affil{5}{Boston University, Boston, MA 02115 USA},
Walter Quattrociocchi \affil{1},
\affil{6}{corresponding author walter.quattrociocchi@imtlucca.it}
}



\significancetext{SIGNIFICANCE: Using a massive quantitative analysis of Facebook, we
show that information related to very specific narratives -- conspiracy
theories and scientific news -- generates homogeneous and polarized
communities that have similar information consumption patterns.  To
account for these features we derive a data-driven percolation model
of rumor spreading that demonstrates that homogeneity and polarization
are the main determinants for predicting cascade size.}

\maketitle

\begin{article}
	
\begin{abstract}
		
The wide availability of user-provided content in online social media
facilitates the aggregation of people around common interests,
worldviews, and narratives. Despite the enthusiastic rhetoric on the
part of some that this process generates ``collective intelligence'',
the WWW also allows the rapid dissemination of unsubstantiated
conspiracy theories that often elicite rapid, large, but naive social
responses such as the recent case of Jade Helm 15 -- where a simple military exercise turned out to be perceived as the beginning of the civil war in the US. 
We study how Facebook users consume information related to two different kinds of narrative:
scientific and conspiracy news. We find that although consumers
of scientific and conspiracy stories present similar consumption
patterns with respect to content, the sizes of the spreading cascades
differ. Homogeneity appears to be the primary driver for the diffusion
of contents, but each echo chamber has its own cascade dynamics. 
To mimic these dynamics, we introduce a data-driven percolation model on
signed networks.
		
\end{abstract}

\keywords{misinformation | rumor spreading | collective narratives | crowd dynamics | online social media }
	
\dropcap{T}he massive diffusion of socio-technical systems and
microblogging platforms on the WWW creates a direct path from producers
to consumers of content, i.e., allows disintermediation, and changes the
way users become informed, debate, and form their opinions
\cite{brown2007word,Richard2004,QuattrociocchiCL11,Quattrociocchi2014,Kumar2010}. This disintermediated environment can foster confusion about causation,
and thus encourage speculation, rumors, and mistrust
\cite{sunstein2009conspiracy}. In 2011 a blogger claimed that global
warming was a fraud designed to diminish liberty and weaken democracy
\cite{Forbes2011}. Misinformation about the Ebola epidemic has caused
confusion among health-care workers \cite{RumorEbola3}.  
Recent research \cite{bessi2014science,mocanu2014,bessi2014economy} has shown that increasing the exposure of users to unsubstantiated rumors increases
their tendency to be credulous.
	
According to Ref.~\cite{furedi2006culture}, beliefs formation and
revision is influenced by the way communities attempt to make sense to
events or facts.  Such a phenomenon is particularly evident on the WWW
where users, embedded in homogeneous clusters
\cite{aiello2012friendship,gu2014research,bessi2014viral}, process
information through a shared system of meaning
\cite{bessi2014science,mocanu2014}.
	
Here we analyze the cascade dynamics of Facebook users when the content
is (i) conspiracy theories and (ii) scientific information. Conspiracy
theories simplify causation, reduce the complexity of reality, and
contain uncertainty
\cite{byford2011conspiracy,finerumor,hogg2011extremism}.  Scientific
information disseminates scientific advances and exhibits the process of
scientific thinking. The main difference between the two is content
verifiability. The generators of scientific information and their data,
methods, and outcomes are readily identifiable and available. The
origins of conspiracy theories are often unknown and the content of the
theories is strongly disengaged from mainstream society and sharply
divergent from recommended practices \cite{betsch2013debunking}, e.g.,
belief that vaccinations cause autism.
	
Massive digital misinformation is becoming pervasive in online social
media to the extent that it has been listed by the World Economic Forum
(WEF) as one of the main threats to our society \cite{Davos13}. To counteract this trend, algorithmic-driven solutions have been proposed
\cite{qazvinian2011rumor,ciampaglia2015computational,resnick2014rumorlens,gupta2014tweetcred,almansour2014model,ratkiewicz2011detecting},
e.g., Google \cite{dong2015knowledge} is developing a {\em trustworthiness score\/} to rank the results of queries. Similarly, Facebook has proposed a community-driven approach where users can flag false contents to correct the news-feed algorithm.  This issue is	controversial, however, because it raises fears that the free	circulation of content may be threatened and the proposed algorithms not	be accurate or effective \cite{bessi2014science,mocanu2014,Nyhan01042014}. Often conspiracists
will denounce attempts to debunk false information, e.g., the link
between vaccination and autism, as acts of misinformation.
	
Whether a claim (either substantiated or not) is accepted by an
individual is strongly influenced by social norms and the claim's
coherence with the individual's belief system
\cite{Zhu2010,Loftus2011}. Despite some enthusiastic claims about the
growth of a {\em collective intelligence\/} \cite{surowiecki2005wisdom},
many mechanisms animate the flow of false information that generates
false beliefs in an individual, which, once adopted, are rarely
corrected \cite{Garrett2013,Meade2002,koriat2000,Ayers98}.
	
We use quantitative analysis to show that homogeneity is the primary
driver of content diffusion and generates the formation of homogeneous,
polarized clusters, i.e., ``echo chambers'' \cite{bessi2014science,mocanu2014,sunstein2001echo,garrett2009echo}. 
We also find that although consumers of scientific information and conspiracy theories exhibit similar consumption patterns with respect to content, the cascade	patterns of the two differ.  Homogeneity appears to be the preferential driver for the diffusion of content, yet each echo chamber has its own	cascade dynamics.  

The paper is structured as follows. First we provide the preliminary
definitions and details concerning data collection. We then do a
comparative analysis and characterize the statistical signatures of the
cascades of the different kinds of content. Finally, we introduce
a data-driven model that replicates the analyzed cascade
dynamics.
	
\section*{Methods}
	
\subsection*{Ethics Statement.}

The data collection process has been carried out using the Facebook
Graph API \cite{fb_graph_api}, which is publicly available. For the
analysis (according to the specification settings of the API) we only
used publicly available data (thus users with privacy restrictions are
not included in the dataset). The pages from which we download data are
public Facebook entities and can be accessed by anyone. User content
contributing to these pages is also public unless the user's privacy
settings specify otherwise, and in that case it is not available to us.
	
\subsection*{Data collection.}
	
Debate about social issues continues to expand across the Web, and
unprecedented social phenomena such as the massive recruitment of people
around common interests, ideas, and political visions are
emerging. Using the approach described in Ref.~\cite{bessi2014science},
we define the space of our investigation with the support of diverse
Facebook groups that are active in the debunking of conspiracy theories.

The resulting dataset is composed of 67 public pages divided between
conspiracy and science news.  A second set, composed of two
troll pages, is used as a benchmark to fit our data-driven model.  The
first category (conspiracy theories) includes the pages that disseminate
alternative, controversial information, often lacking supporting
evidence and frequently advancing conspiracy theories.  The second
category (science news) includes the pages that disseminate scientific
information. The third category (trolls) includes those pages that
intentionally disseminate sarcastic false information on the Web.
	
For the three sets of pages we download all the posts (and their
respective user interactions) across a five-year timespan (2010 to
2014).  We perform the data collection process by using the Facebook
Graph API \cite{fb_graph_api}, which is publicly available and
accessible through any personal Facebook user account.  The exact
breakdown of the data is presented in the Supporting 
Information (SI) Section~1.

\subsection*{Preliminaries and Definitions.}

A tree is an undirected simple graph that is connected and has no simple
cycles. An oriented tree is a directed acyclic
graph whose underlying undirected graph is a tree. A sharing tree in the
context of our research is
an oriented tree made up of the successive sharing of a news item through the
Facebook system. The root of the sharing tree is the node that performs
the first temporal share.  We define the size of the sharing tree as the
number of nodes (and hence the number of news sharers) in the
tree and the height of the sharing tree as the maximum path length
distant
 from the root. 
 
We define the user polarization $\sigma= 2\varrho -1$, where $0\leq \varrho \leq 1$ is the fraction of ``Likes'' a user executes on
conspiracy related content, and hence $-1\leq
\sigma\leq 1$. From user polarization, we define the
edge homogeneity, for any edge $e_{ij}$ between
nodes $i$ and $j$, as
			$$
			\sigma_{ij} = \sigma_i\sigma_j, 
			$$ 
with $-1 \leq \sigma_{ij} \leq 1$.  
Edge homogeneity reflects the similarity level between the polarization of the two sharing nodes. A link in the sharing tree is homogeneous if its edge homogeneity is positive, otherwise it is non homogeneous.
We then define a sharing path to be any path from the root to one of the leaves of the sharing tree. A homogeneous path is a sharing path for which the edge homogeneity of each edge is positive, i.e., a sharing path whose edges are all homogeneous links.

\paragraph{Wald Test.} We use the Wald test to compare
the scaling parameters of two power law distributions. We define it as
\begin{eqnarray*}
		H_0 : \hat{\alpha_1} = \hat{\alpha_2}\\ H_1 :
                \hat{\alpha_1} \neq \hat{\alpha_2}
\end{eqnarray*}
where $\hat{\alpha_1}$ and $\hat{\alpha_2}$ are the estimated scaling
parameters. The Wald statistic:
$$
		W = \frac{(\hat{\alpha_1} -\hat{\alpha_2})^2}{Var(\hat{\alpha_1})},
$$
follows a $\chi^2$ distribution with one degree of freedom.  We reject
the null hypothesis $H_0$ and conclude that there is a significant
difference between the two scaling parameters if the $p$-value of $W$ is
below a given significance level $\alpha$.

\paragraph{Kolmogorov-Smirnov Test.} We use the Kolmogorov-Smirnov test 
to compare the empirical distribution functions of two samples. The
Kolmogorov-Smirnov statistic for two given cumulative distribution
functions $F_1(x)$ and $F_2(x)$ is 
		$$
	    D = \sup_x{|F_1(x) -F_2(x)|},
		$$

which measures the maximum punctual distance between the two sample
distributions. If $D$ is bigger than a given critical value
$D_{\alpha}$\footnote{The critical value $D_{\alpha}$ depends on the
  sample sizes and on the considered significance level $\alpha$, it can
  be computed as
			$$
			D_{\alpha} = c(\alpha)\sqrt{\frac{n_1 + n_2}{n_1n_2}},
			$$
where $n_1$ and $n_2$ are the respective sample sizes and $c(\alpha)$ is
a fixed value associated with the significance level $\alpha$.} we
reject the null hypothesis $H_0 : F_1(x) = F_2(x)$ and conclude that
there is a significant difference between the two sample distributions.

\section*{Results and discussion}

\subsection*{Anatomy of Cascades.}

We begin our analysis by characterizing the statistical signature of
cascades as they relate to information type. We analyze the three
types---science news, conspiracy rumors, and trolling---and find that
size and maximum degree are power-law distributed for all three. The
maximum cascade size values are 952 for science news, 2422 for
conspiracy news, and 3945 for trolling, and the estimated exponents
for the power law distributions are 2.21 for science news, 2.47 for
conspiracy theories, and 2.44 for trolling.  Tree height values range
from 1 to 5, with a maximum height of 5 for science news and conspiracy
theories and a maximum height of 4 for trolling. For further information
see SI Section~2.1.

\begin{figure}
	\includegraphics[width=.35\textwidth]{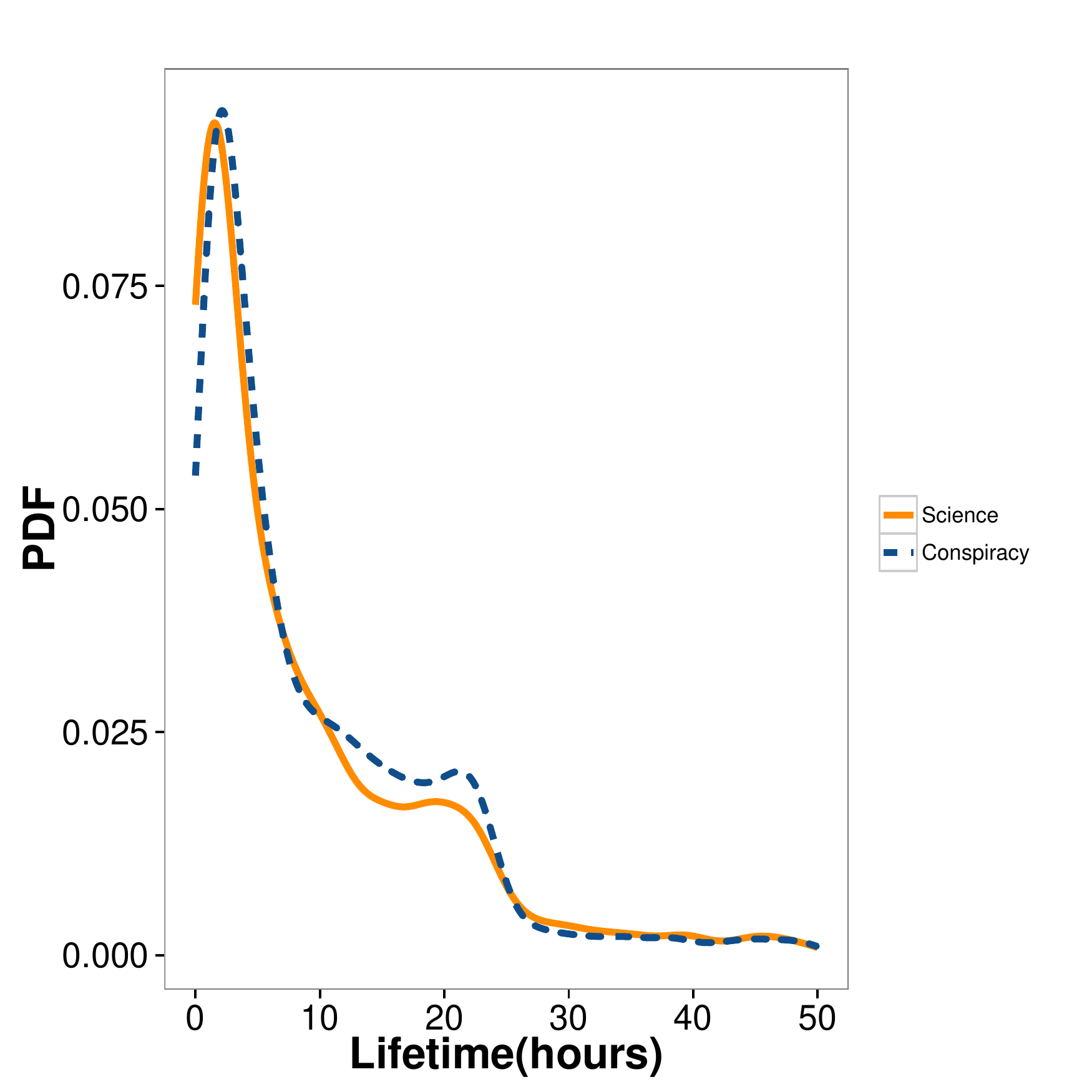}
	
	\caption{Probability density function
		(PDF) of Lifetime computed on science news and conspiracy
		theories, where the lifetime is here computed as the
		temporal distance (in hours) between the first and
		last share of a post. Both categories show a similar
		behavior, with a peak in the first two hours and
		another around 20 hours. }\label{fig:lf_tot}
\end{figure}

Figure~\ref{fig:lf_tot} shows the probability density function (PDF) of
the cascade lifetime (using hours as time units) for science and
conspiracy. We compute the lifetime as the length of time
between the first user and the last user sharing a post.  In both
categories we find a first peak at approximately 1--2 hours and a second
at approximately 20 hours, indicating that the temporal sharing patterns
are similar irrespective of the difference in topic. We also find that a
significant percentage of the information diffuses rapidly (24.42\% of
the science news and 20.76\% of the conspiracy rumors diffuse in less
than two hours, and 39.45\% of science news and 40.78\% of conspiracy
theories in less than five hours).  Only 26.82\% of the diffusion of
science news and 17.79\% of conspiracy lasts more than one day.
Kolmogorov-Smirnov test made us reject the hypothesis  $H_0$
that the two distributions are equal.

\begin{figure}
	\includegraphics[width=0.5\textwidth]{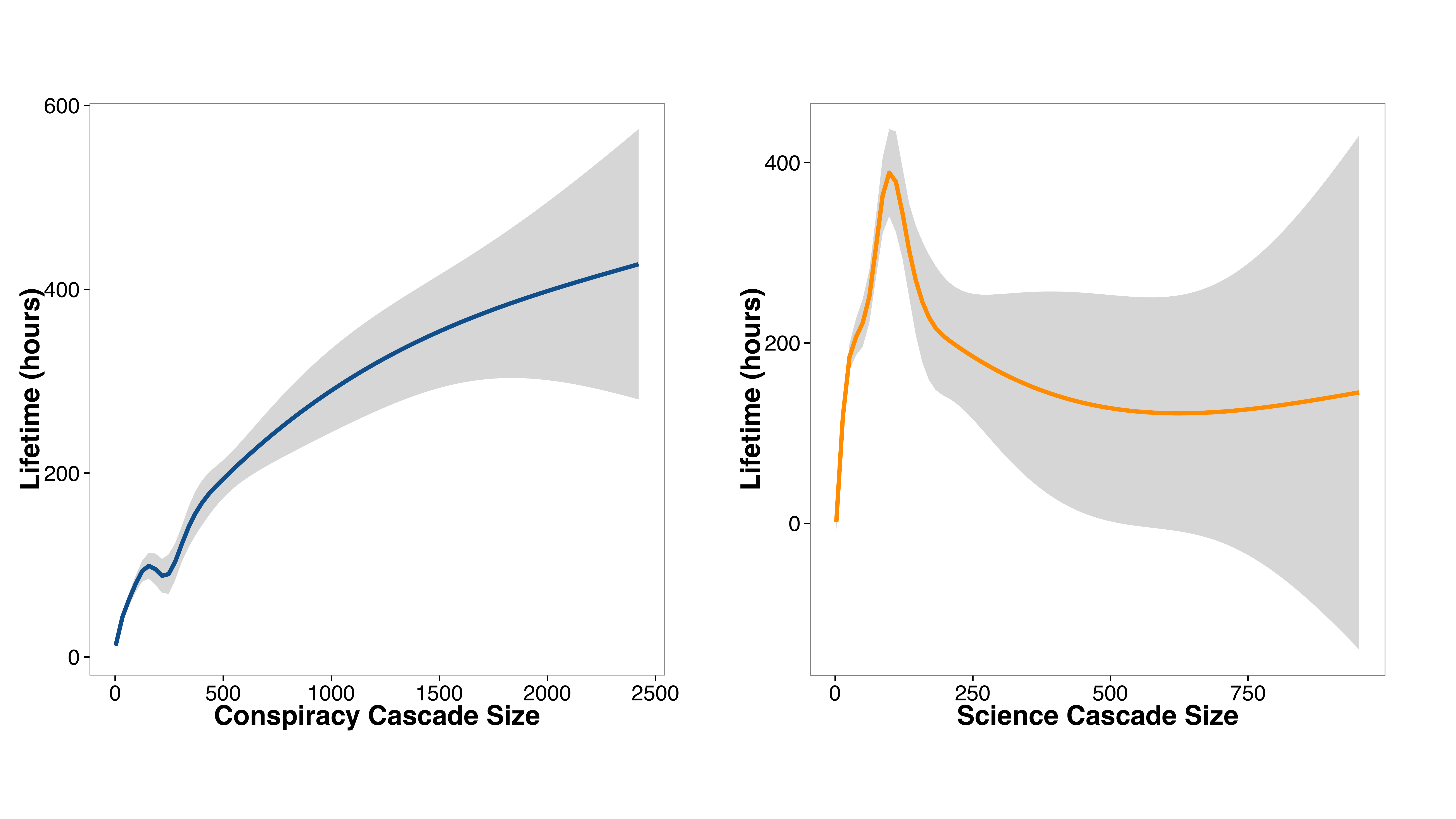}
	\caption{Lifetime as a function of the cascade
		size for conspiracy news (left) and science news (right). We
		note a contents-driven differentiation in the sharing
		patterns. For conspiracy the lifetime grows with the
		size, while for science news there is a peak in the lifetime
		around a value of the size equal to 200, and a higher
		variability in the lifetime for larger cascades.}\label{fig:lf_size}      
\end{figure}

Figure~\ref{fig:lf_size} shows lifetime as a function of cascade size.
For science news we have a peak in the lifetime corresponding to a
cascade size value of $\approx 200$, and higher cascade size values
correspond to high lifetime variability.  
For conspiracy related content the lifetime increases with cascade size.

These results suggest that news assimilation differs according to
category. Science news is usually assimilated, i.e., it reaches a higher
level of diffusion, quickly, and a longer lifetime does not correspond
to a higher level of interest.  
Conversely, conspiracy rumors are assimilated more
slowly and show a positive relation between lifetime and size. 
For both science and conspiracy news, we compute size as a function
of lifetime and confirm that differentiation in the sharing patterns is
content-driven, and that for conspiracy there is a positive
relation between size and lifetime. For a more detailed explanation, see
SI Section~2.1.

\subsection*{Homogeneous Clusters.}

We next examine the social determinants that drive sharing patterns and
we focus on the role of homogeneity in friendship networks.

\begin{figure}
	\includegraphics[width=0.35\textwidth]{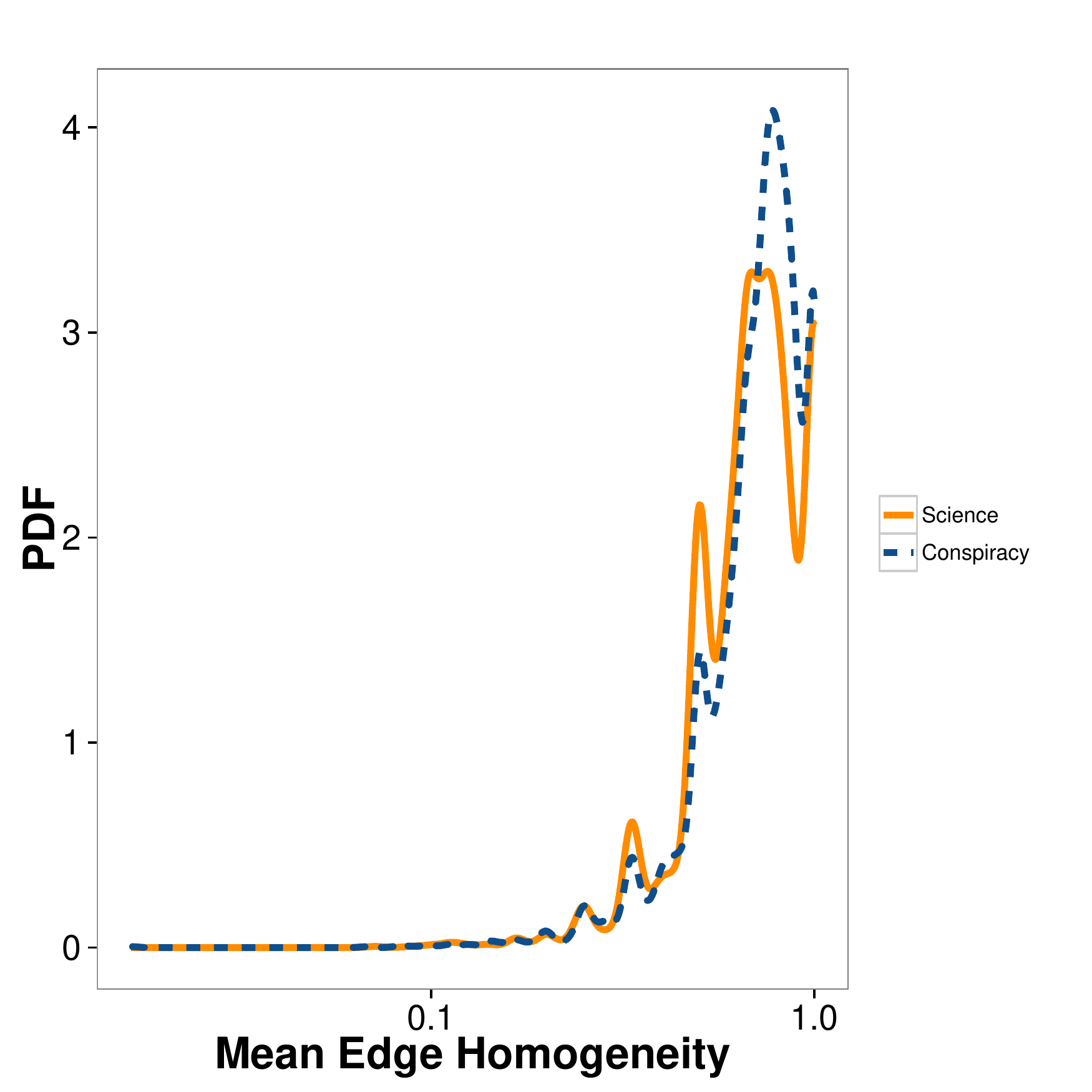}
	\caption{Mean edge homogeneity for science  (solid orange) and conspiracy  (dashed blue) news. The mean value of edge homogeneity on the whole sharing cascades is always greater or equal to zero.}\label{fig:polarization}      
\end{figure}

Figure~\ref{fig:polarization} shows the PDF of the mean edge homogeneity, computed for all cascades of science news and conspiracy theories. It shows that there are homogeneous links
between consecutively sharing users. In particular, the average edge homogeneity
value of the entire sharing cascade is always greater or equal to zero, indicating that either the information transmission occurs inside homogeneous clusters in which all links are homogeneous or it occurs inside mixed neighborhoods in which the balance between homogeneous and non homogeneous links is favorable towards the former ones. However, the probability of close to zero mean edge homogeneity is really small.  

\begin{figure}
	\includegraphics[width=0.35\textwidth]{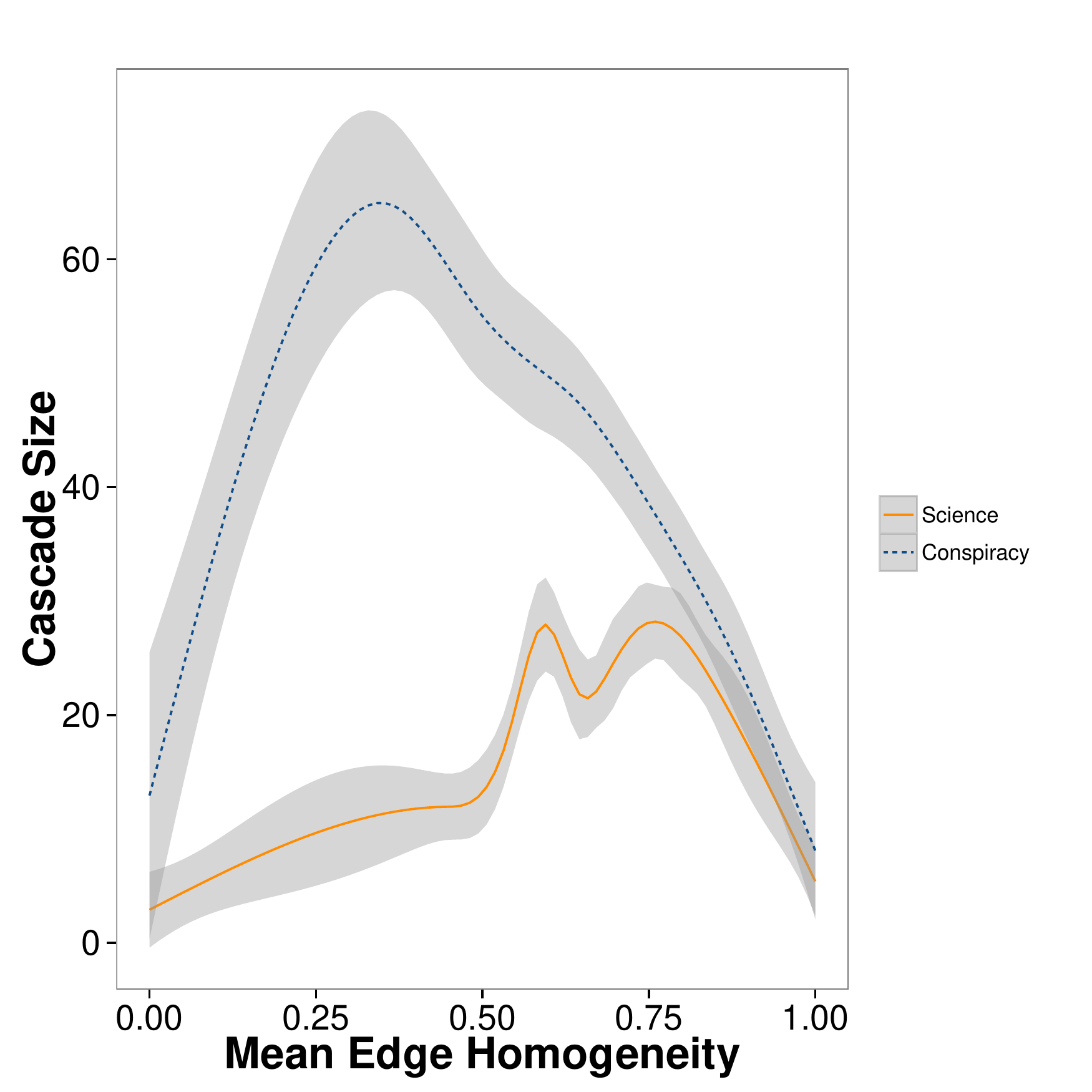}
	\caption{Cascade size as a function of mean edge homogeneity for science  (solid orange) and conspiracy (dashed blue) news.}\label{fig:size_polarization}      
\end{figure}

To further characterize the role of homogeneity in shaping sharing
cascades, we compute cascade size as a function of mean edge homogeneity for both
science and conspiracy news, see Figure~\ref{fig:size_polarization}. In science news, higher levels of mean edge homogeneity in the
interval (0.5, 0.8) correspond to larger cascades, but in conspiracy
theories lower levels of mean edge homogeneity ($\sim 0.25$) correspond to larger
cascades.
Notice that, although viral patterns related to distinct contents differ, homogeneity is clearly the driver of information diffusion. In other words, different contents generate different echo chambers, characterized by the high level of homogeneity inside them.

The probability density function (PDF) of the edge homogeneity, computed for science and conspiracy news as well as the two taken together---both in the unconditional case and in the conditional case (in the event that the user that made the first share in the couple has a positive or negative polarization)---confirms the roughly null probability of a negative edge homogeneity (see SI Section~2.1).


We record the CCDF of the number of all sharing paths\footnote{Recall that a sharing path is here defined as any path from the root to one of the leaves of the sharing tree. A homogeneous path is a sharing path for which the edge homogeneity of each edge is positive} on each tree 
compared with the CCDF of the number of homogeneous paths for science and conspiracy news, and the two together.  A Kolmogorov-Smirnov test and Q-Q plots confirm that for all three pairs of distributions considered there is no significant statistical
difference (see SI Section~2.2 for a more detailed analysis).
In SI Section~2.2 we report also the frequency of maximum length for all sharing paths and homogeneous paths, for both categories of content.
 
We confirm the pervasiveness of homogeneous paths, but we also find homogeneous paths in which there is a shift of $-1$ in the path length (with respect to the total path length $k$). 
Notice that the first publisher of a news is generally a page, hence the $(k-1)$-homogeneous paths are due to a discordant sharing in
the first step (i.e., when the product of the first sharer's user polarization
and the sharer page category is negative). 

Cascade lifetimes of science and conspiracy
news exhibit a probability peak in the first two hours, and that in
the following hours they rapidly decrease.  Despite the similar
consumption patterns, cascade lifetime expressed as a function of
cascade size differs greatly for the different content sets. 
The PDF of the mean edge homogeneity indicates that there is homogeneity in the linking step
of sharing cascades.  
The distribution of the number of total and homogeneous sharing paths are very similar for both content categories.

Viral patterns related to contents belonging to different narratives differ, but homogeneity is
clearly the driver of content diffusion.

\subsection*{The Model.}

We now introduce a percolation model of rumor spreading to account for
homogeneity and polarization.  We consider $n$ users connected by a
small-world network \cite{watts1998collective}. 
The model parameter space varies on a rewiring
probability $r$, mimicking the network density, and a news set of size $m$.

Every node has an opinion $\omega_i$, $i\in[1,n]$ uniformly distributed
in $[0,1]$. Every news item has a fitness (degree of interest)
$\vartheta_j,\,j\in[1,m]$ uniformly distributed in $[0,1]$.  
At each step the news items are diffused and initially shared by a
group of first sharers.  After the first step, the news recursively
passes to the neighborhoods of previous step sharers, e.g., those of the
first sharers during the second step. 
If a friend of the previous step sharers has an opinion close to the fitness of the news, then she shares the news again. 

In particular, when
$$
{|\omega_i -\vartheta_j|\leq\delta},
$$
user $i$ shares news $j$; $\delta$ is the sharing threshold.

Because $\delta$ by itself cannot capture the homogeneous clusters
observed in the data, we model the connectivity pattern as a signed network \cite{Quattrociocchi2014,leskovec2010signed} considering different fractions of homogeneous
links and hence restricting diffusion of news only to homogeneous
links. 
We define $\phi_{HL}$ as the fraction of homogeneous links in the network, $M$ as the number of total links, and $n_h$ as the number of homogeneous links, thus we have:
$$
\phi_{HL} = \frac{n_h}{M},\,0\leq n_h\leq M.
$$
Notice that $0\leq \phi_{HL}\leq 1$ and that $1 - \phi_{HL}$, the fraction of non homogeneous links, is complementary to $\phi_{HL}$. In particular, we can reduce the parameters space to $\phi_{HL}\in[0.5,1]$  as we would restrict our attention to either one of the two complementary clusters.

The model can be seen as a branching process where the sharing
threshold $\delta$ and neighborhood dimension $z$ are the key parameters.
More formally, let the fitness $\theta_{j}$ of the $j^{th}$ news and the
opinion $\omega_{i}$ of a the $i^{th}$ user be uniformly i.i.d. between
$[0,1]$.  Then the probability $p$ that a user $i$ shares a post $j$ is
defined by a probability $p=\min(1,\theta + \delta) -
\max(0,\theta-\delta)\approx2\delta$, since $\theta$ and $\omega$ are
uniformly i.i.d.  In general, if $\omega$ and $\theta$ have
distributions $f(\omega)$ and $f(\theta)$, then $p$ will depend on
$\theta$,
\[
p_{\theta}=f\left(\theta\right)\int_{\max\left(0,\theta-
	\delta\right)}^{\min\left(1,\theta+
	\delta\right)}f\left(\omega\right)d\omega.  
\]
If we are on a tree of degree $z$ (or on a sparse lattice of degree
$z+1$), the average number of sharers (the branching ratio) is defined
by
\[
\mu=zp\approx2\delta\, z
\]
with a critical cascade size $S=\left(1-\mu\right)^{-1}$. If we assume
that the distribution of the number $m$ of the first sharers is
$f\left(m\right)$, then the average cascade size is
\[
S=\sum_{m}f\left(m\right)m\left(1-\mu\right)^{-1}=
\frac{\left\langle m\right\rangle _{f}}{1-\mu}
\approx\frac{\left\langle m\right\rangle _{f}}{1-2\delta z}
\]
where $\left\langle \ldots\right\rangle _{f}=\sum_{m}\ldots f\left(m\right)$
is the average with respect to $f$.
In the simulations we fixed  neighborhood dimension $z = 8$ since the branching ratio $\mu$ depends upon the product of $z$ and $\delta$ and, without loss of generality, we can consider the variation of just one of them.

If we allow a probability $q$ that a neighbor of a user has a
different polarization, then the branching ratio becomes
$\mu=z\left(1-q\right)p$.  If a lattice has a degree distribution
$d\left(k\right)$ ($k=z+1$), we can then assume a usual percolation
process that provides a critical branching ratio and that is linear in
$\left\langle k^{2}\right\rangle _{d}/\left\langle k\right\rangle _{d}$
($\mu\approx\left(1-q\right)p\left\langle z^{2}\right\rangle
/\left\langle z\right\rangle $).

\subsection*{Simulation Results.}

We explore the model parameters space using $n = 5,000$ nodes and $m =
1,000$ news items with the number of first sharers distributed as an (i)
inverse Gaussian, (ii) log normal, (iii) Poisson, (iv) uniform
distribution, and as the real data distribution (from the science  and
conspiracy news sample). Parameters are chosen to fit the real data
distribution (for details see SI Section~3.1,~3.2). In Table~\ref{tab1} we show a summary of relevant statistics (min value, first quantile, median, mean, third quantile, and max value) to compare the real data first sharers distribution with the fitted distributions\footnote{For details on the parameters of the fitted distributions used see SI Section~3.2.}. The inverse Gaussian ($IG$), 
shows the best fit for the distribution of first sharers with respect to all the considered statistics. 

Along with the first sharers distribution, we vary the sharing threshold $\delta$ in the interval $[0.01, 0.05]$ and the fraction of homogeneous links $\phi_{HL}$ in the interval $[0.5, 1]$.  To avoid biases induced by statistical
fluctuations in the stochastic process, each point of the parameter
space is averaged over 100 iterations.  $\phi_{HL}\sim 0.5$ provides
a good estimate of real data values.  In particular, consistently with the division of in two echo chambers (science and conspiracy), 
the network is divided into two clusters in which news items remain
inside and are transmitted solely within each community's echo chamber
(see SI Section~3.2 for the details of the simulation
results).

In addition to the science and conspiracy content sharing trees, we
downloaded a set of 1,072 sharing trees of intentionally false
information from troll pages. Frequently troll information, e.g.,
parodies of conspiracy theories such as chem-trails containing the
active principle of Viagra, is picked up by habitual conspiracy theory
consumers. In SI Section~3.2 we report the same information as Table~\ref{tab1} for trolling category. Also in this case we notice that the best fit is obtained by the inverse Gaussian distribution. 

We computed the mean and standard deviation of 
size and height of all trolling sharing trees, and reproduced the data 
using our model\footnote{Note that the real data values for the mean (and standard deviation) of size and height on the troll posts are respectively: $23.54\, (122.32)$ and $1.78\,(0.73)$.}. We used fixed 
parameters from trolling messages sample (the number of nodes in the system and the number of news items) and varied the fraction of homogeneous links $\phi_{HL}$, the rewiring probability $r$, and sharing threshold $\delta$. See SI Section~3.2 for the distribution of first sharers used and for additional simulation results of the fit on trolling messages.

\begin{figure}
	\includegraphics[width=0.5\textwidth]{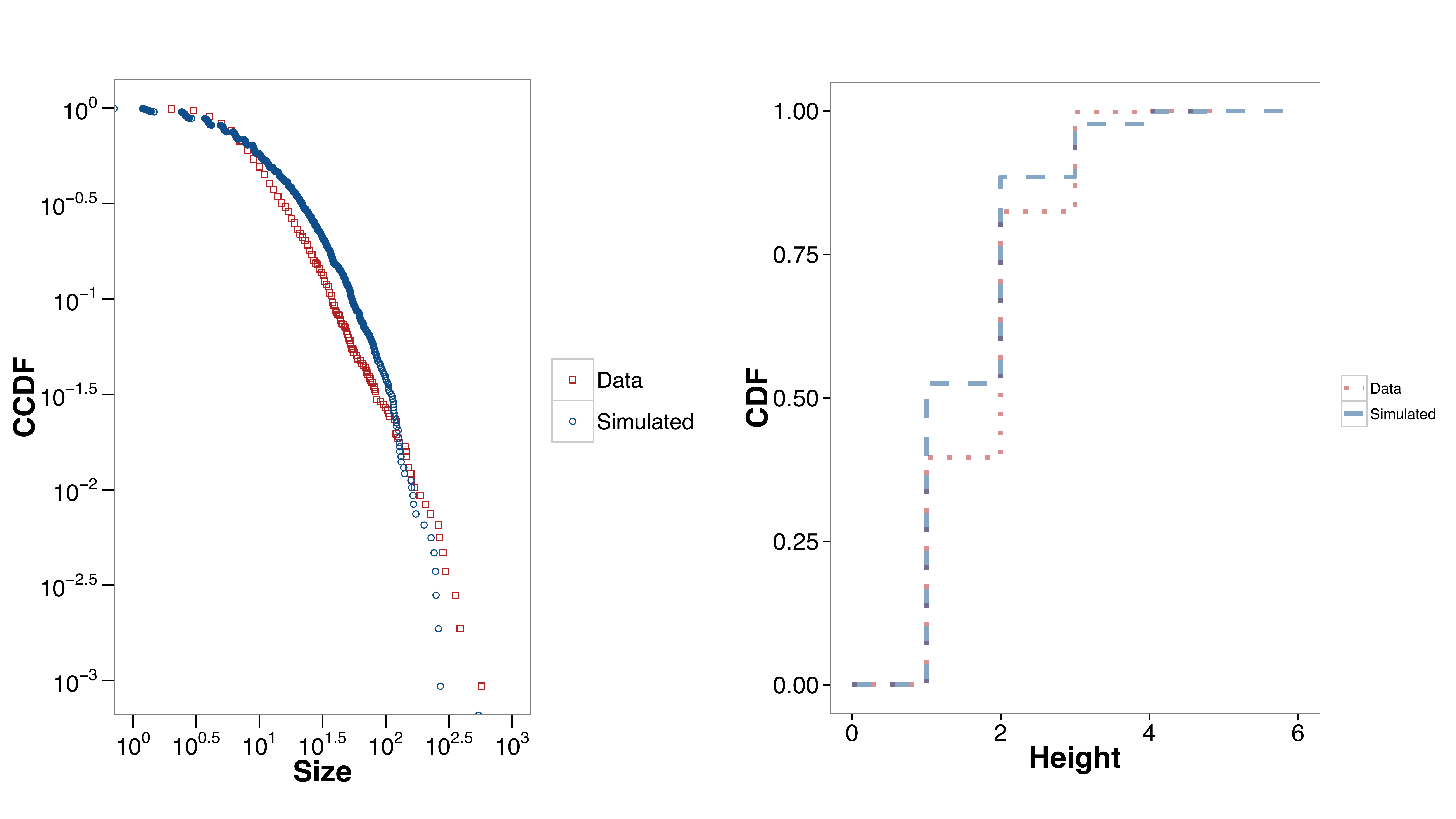}
	\caption{Complementary cumulative distribution
		function (CCDF) of size  (left) and cumulative distribution function (CDF) of height (right) for the best parameters
		combination that fits troll data values, $(\phi_{HL}, r,
		\delta) = (0.56, 0.01, 0.015)$, and first sharers
		distributed as $IG(18.73, 9.63)$. We note that it is indeed a good fit of trolling data.}
	\label{fig5}      
\end{figure}

We simulated the model dynamics with the best combination of parameters obtained from the simulations  and the number first sharers distributed as an inverse Gaussian, figure~\ref{fig5} shows the CCDF of size and the CDF of height. A summary of relevant statistics (min value, first quantile, median, mean, third quantile, and max value) to compare the real data size and height distributions with the fitted ones is reported in SI Section~3.2. We notice that the fit is good for all the statistics, with the exception of min and max value of size. For the min value, the presence of a zero is due to the fact that the inverse Gaussian is a real valued distribution function and in the simulations we considered the integer part of the number of first sharers, thus producing a number of never shared pieces of information. On the other hand, the high difference in the max value is probably due to the long tail of the data size distribution.  

We find that the inverse Gaussian is the distribution that best fits the
data results both for science and conspiracy news, and for troll messages.  For this reason, we performed one more simulation using the inverse Gaussian as distribution of the number of first sharers, 1,072 news items, 16,889 users, and the best parameters combination obtained in the simulations
\footnote{The best parameters combinations is $	\phi_{HL} = 0.56, \, r = 0.01, \, \delta = 0.015$, in this case we have a mean size equal to $23.42\,(33.43)$ and a mean height $1.28\,(0.88)$, and it is indeed a good approximation, see Section~3.2.}.
The CCDF of size and the CDF of height for the above parameters
combination, as well as basic statistics considered, fit the real data ones from the trolling category.

\section*{Conclusions}

Digital misinformation has become so pervasive in online social media
that it has been listed by the World Economic Forum (WEF) as one of the
main threats to human society. Whether a news item, either
substantiated or not, is accepted as true by a user may be strongly
affected by social norms or by how much it coheres with the user's
system of beliefs \cite{Zhu2010,Loftus2011}. Despite enthusiastic claims
that social media is generating a vast ``collective intelligence''
available to all \cite{surowiecki2005wisdom}, many mechanisms cause
false information to gain acceptance, which in turn generate false
beliefs that, once adopted by an individual, are highly resistant to
correction \cite{Garrett2013,Meade2002,koriat2000,Ayers98}. Using extensive
quantitative analysis we show that social homogeneity is the primary
driver of content diffusion, and one frequent result is the formation of
homogeneous, polarized clusters (often called ``echo chambers'').  We
also find that although consumers of science news and conspiracy
theories show similar consumption patterns with respect to content,
their cascades differ. Social homogeneity appears to be the primary
driver of content diffusion, and each echo chamber has its own cascade
dynamics.  To mimic these dynamics, we introduce a data-driven
percolation model of signed networks, i.e., networks composed of signed
edges.  Our analysis shows that for science and conspiracy news
a cascade's lifetime has a probability peak in the first two hours
followed by a rapid decrease. Although the consumption patterns are
similar, cascade lifetime as a function of the size differs greatly.
The PDF of the mean edge homogeneity indicates that
homogeneity is present in the linking step of sharing cascades.  The
distribution of the number of total sharing paths and homogeneous
sharing paths are similar in both content categories.  Viral patterns
related to distinct contents are different but homogeneity drives
content diffusion. We simulate our data-driven percolation model by
fixing the number of users and news items downloaded from troll pages
and varying the other parameters. 
We compare the simulated results with the data and find a high level of similarity.


\begin{acknowledgments}
Funding for this work was provided by the EU FET project MULTIPLEX, no. 317532, SIMPOL, no. 610704, the FET project DOLFINS 640772 (H2020), SoBigData 654024 (H2020), and CoeGSS 676547 (H2020).
The funders had no role in study design, data collection and analysis, decision to publish, or preparation of the manuscript.  Special thanks go to Delia
Mocanu,"Protesi di Protesi di Complotto", "Che vuol dire reale", "La
menzogna diventa verita e passa alla storia", "Simply Humans",
"Semplicemente me", Salvatore Previti, Brain Keegan, Dino Ballerini,
Elio Gabalo and "The rooster on the trash" for their precious
suggestions and discussions.

\end{acknowledgments}

\bibliographystyle{unsrt}
\bibliography{biblio_final}
\end{article}


\begin{table}[H]
	\centering
	{\large{
			\begin{tabular}{ c | c | c | c | c }
			\hline
			& Data & IG & LN   & Poi \\ 
			\hline
			Min & \textit{1}&\textbf{0.36}& 0.10& 20  \\
			1st Qu.& \textit{5}&\textbf{4.16}& 3.16& 35\\
			Median  & \textit{10}&\textbf{10.45}& 6.99& 39 \\
			Mean& \textit{39.34}&\textbf{39.28}& 13.04& 39.24\\
			3rd Qu. & \textit{27}&\textbf{31.59}& 14.85 &  43 \\
			Max & \textit{3033}&\textbf{1814}& 486.10&  \,\,\,\,66
			\caption{Summary of relevant statistics for the first sharers distributions.}\label{tab1}	
			\end{tabular}
		
		}}
\end{table}

\end{document}